\documentstyle[12pt,epsf]{article}
\def\hybrid{\topmargin 0pt      \oddsidemargin 0pt
	\headheight 0pt \headsep 0pt
	\textheight 9in         
	\textwidth 6.25in       
	\marginparwidth .875in
	\parskip 5pt plus 1pt   \jot = 1.5ex}

\catcode`\@=11
\def\marginnote#1{}
\newcount\hour
\newcount\minute
\newtoks\amorpm
\hour=\time\divide\hour by60
\minute=\time{\multiply\hour by60 \global\advance\minute by-\hour}
\edef\standardtime{{\ifnum\hour<12 \global\amorpm={am}%
	\else\global\amorpm={pm}\advance\hour by-12 \fi
	\ifnum\hour=0 \hour=12 \fi
	\number\hour:\ifnum\minute<10 0\fi\number\minute\the\amorpm}}
\edef\militarytime{\number\hour:\ifnum\minute<10 0\fi\number\minute}

\def\draftlabel#1{{\@bsphack\if@filesw {\let\thepage\relax
   \xdef\@gtempa{\write\@auxout{\string
      \newlabel{#1}{{\@currentlabel}{\thepage}}}}}\@gtempa
   \if@nobreak \ifvmode\nobreak\fi\fi\fi\@esphack}
	\gdef\@eqnlabel{#1}}
\def\@eqnlabel{}
\def\@vacuum{}
\def\draftmarginnote#1{\marginpar{\raggedright\scriptsize\tt#1}}

\def\draft{\oddsidemargin -.5truein
	\def\@oddfoot{\sl preliminary draft \hfil
	\rm\thepage\hfil\sl\today\quad\militarytime}
	\let\@evenfoot\@oddfoot \overfullrule 3pt
	\let\label=\draftlabel
	\let\marginnote=\draftmarginnote
   \def\@eqnnum{(\theequation)\rlap{\kern\marginparsep\tt\@eqnlabel}%
\global\let\@eqnlabel\@vacuum}  }

\def\numberbysection{\@addtoreset{equation}{section}
	\def\theequation{\thesection.\arabic{equation}}}

\def\underline#1{\relax\ifmmode\@@underline#1\else
	$\@@underline{\hbox{#1}}$\relax\fi}

\def\titlepage{\@restonecolfalse\if@twocolumn\@restonecoltrue\onecolumn
     \else \newpage \fi \thispagestyle{empty}\c@page\z@
	\def\thefootnote{\fnsymbol{footnote}} }

\def\endtitlepage{\if@restonecol\twocolumn \else  \fi
	\def\thefootnote{\arabic{footnote}}
	\setcounter{footnote}{0}}  
\catcode`@=12
\relax

\def\ve{\varepsilon}
\def\nn{\nonumber}
\def\beq{\begin{equation}}
\def\eeq{\end{equation}}
\def\bea{\begin{eqnarray}}
\def\eea{\end{eqnarray}}
\def\ep{\epsilon}

\relax
\numberbysection
\hybrid

\begin{document}
\begin{titlepage}
\begin{center}
April~1996 \hfill    PAR--LPTHE 95/63 \\
$~$ \hfill hep-th/9512087\\
[.5in]
{\large\bf Randomly coupled minimal models}\\[.5in]
        {\bf Vladimir Dotsenko\footnote{Also at the Landau Institute for
Theoretical Physics, Moscow}, Marco Picco and Pierre
Pujol} \\
	{\it LPTHE\/}\footnote{Laboratoire associ\'e No. 280 au CNRS}\\
       \it  Universit\'e Pierre et Marie Curie, PARIS VI\\
       \it Universit\'e Denis Diderot, PARIS VII\\
	Boite 126, Tour 16, 1$^{\it er}$ \'etage \\
	4 place Jussieu\\
	F-75252 Paris CEDEX 05, FRANCE\\
	dotsenko,picco,pujol@lpthe.jussieu.fr
\end{center}

\vskip .5in
\centerline{\bf ABSTRACT}
\begin{quotation}
Using 1-loop renormalisation group equations,
we analyze the effect of randomness on multi-critical
unitary minimal conformal models. We study the case of two
randomly coupled $M_p$ models and found that they flow in two decoupled
$M_{p-1}$ models, in the infra-red limit. This result is then extend to the
case with $M$ randomly coupled $M_p$ models, which will flow toward $M$
decoupled $M_{p-1}$.
\vskip 0.5cm 
\noindent
PACS numbers:  64.60.Ak,64.60.Fr,05.70.Jk

\end{quotation}
\end{titlepage}
\newpage
\section{Introduction}

Recently, many theoretical models have been proposed in studying the effect
of randomness in two-dimensional systems (see for example
\cite{dber}, \cite{cardy}). In a conformal invariant pure model, quenched
impurities introduce a perturbation term which generally breaks conformal
invariance. The effect of a weak disorder can be anticipated by the Harris
criterion \cite{harris}. The relevance of disorder in the critical regime
of a system can be evaluated by power counting. The naive dimensional
analysis of the impurities induced term perturbing the (conformal
invariant) Hamiltonian of the pure system give us the importance of
disorder near the transition point. One of the first results in the case of
uncorrelated quenched disorder was obtained in
\cite{dd1,shalaev,shankar,ludwig} in the context of the Ising model with
random bonds and in \cite{dd2} for the Baxter model.  These cases correspond
to a marginal perturbation of the original Hamiltonian.  The problem becomes
more complicated when quenched impurities induce a relevant perturbation
term. Such an example is found in the $q$-state Potts model with random bonds
\cite{ludwig}, \cite{nous} \cite{marco}. Using the replica method, it was
argued that the 
system reach a non trivial fixed point giving new critical $q$-dependent
exponents.  More recently, in \cite{cardy,sent,domany,pujol} it was shown that
in some cases impurities can drift a first order phase transition to a
continuous one, which is in many cases of Ising-like type. It was pointed
out that, for the 
$q$-state Potts model, the $q$-dependence of exponents for different kinds of
disorder is not obvious.  The absence of non-perturbative results for
uncorrelated disorder forbid us to give a definitive answer.

On the other hand, there exist powerful method in studying conformal field
theories perturbed by some particular operators \cite{zam}. Some exact
results have been obtained by perturbing unitary minimal models. These
models corresponds to multi-critical two-dimensional statistical models
\cite{fried}, \cite{df}. It is interesting to study these models when the
perturbation corresponds to an addition of disorder. This is what we will
do in this paper. The model that we study consists in taking two copies of
such models and add a random energy density depending coupling between
them. Our result is that under particular
conditions, our systems will have a large scale behavior corresponding to
that found in \cite{zam}, namely they flow, in the infrared limit, in a new
minimal model. In section (2) we show how disorder is
implemented in our initial pure systems. In the context of the replica
method we show which perturbing term we have to consider in the effective
Hamiltonian of our replicated system. Then, in section (3), we specialize
in the case of two randomly coupled minimal models. We study the 1-loop
Renormalisation Group (RG) behavior of the system for an arbitrary number
of replica and the quenched case is developed in section (4). In section
(5) we present a generalization to the case with $M$ randomly coupled $M_p$
models with $M \geq 2$ arbitrary. Section (6) is devoted to discussions and
conclusions.
\section{Minimal models perturbed by randomness}
We are interested in conformal invariant multi-critical models perturbed by
randomness. 
Our pure system will then consist in unitary minimal models  $M_{p}$ of
central charge $c=1-{6\over{p(p+1)}}$ \cite{fried}, \cite{df}. 
The conformal dimension of any operator in the conformal grid of the $M_p$
model is
$\Delta_{n',n} = {{\left(pn'-(p+1)n\right)^2 -1}\over{ 4p(p+1)}}$ (physical
dimension is $2\Delta_{n',n}$). For the unitary minimal models
that we consider here, the energy operator correspond to $\ve = \phi_{2,1}$
and has conformal dimension
$\Delta_{2,1} = {1\over4}-{\ep\over4}$ with $\ep = {3\over p+1}$.

In this letter, we will consider a model consisting in $2$ minimal models
$M_{p}$ 
randomly coupled in the following way: if $H_{0,1}$
and $H_{0,2}$ are respectively the Hamiltonians of the two pure models, our
total Hamiltonian is:
\beq
\label{presentation}
H = H_{0,1} + H_{0,2} + \int d^2x ~q(x) ~\ve_1(x) \ve_2(x)
\eeq
where $q(x)$ is a random coupling term between the two models. 
The case of a non-random $q(x)$ was already shown to be integrable and
leads to a massive theory \cite{vaysburd}.
In the context of
the replica method, we take the average of the n$^{th}$ power of the partition
function: 
\beq
\label{z}
\overline{Z^n} = \int \prod_{x} dP[q(x)] Z^n
\eeq
where $dP[q(x)]$ is a symmetric normalized probability distribution for
$q(x)$. 
(\ref{z}) gives naively the effective Hamiltonian:
\beq
\label{qham}
H=\displaystyle\sum_{i=1}^{n} (H_{0,1}^i + H_{0,2}^i)
- \sigma \displaystyle\sum_{i,j=1}^{n} \int d^2x
\left(\ve_{1}^{i} \ve_{2}^{i} \ve_{1}^{j}
\ve_{2}^{j}\right) (x)~+ \cdots
\eeq
Here we just have write the second cumulant of the distribution
$P[q(x)]$. If $p$ is large enough, the dimension of the energy operator will
be close to ${1\over2}$ and naively, higher cumulant terms 
will be irrelevant operators.
The quenched case will be obtained in the limit $n\rightarrow 0$. As a first
stage we will study the model (\ref{qham}) for generic $n$ specializing
then in the quenched case.

We still have to make an important remark, before going to the detailed
calculations. We can see that in both cases, the operator algebra (OA) of
the energy operator contains the operator $ \phi_{3,1}$ and his descendants
\cite{df}~: 
$$
\ve \ve \rightarrow [I] + [\phi_{3,1}]
$$
The conformal dimension of $ \phi_{3,1}$ is given by
$\Delta_{3,1} = 1-{2\ep\over 3}$ and so it is a relevant operator. In
fact, in the 
interaction terms displayed above, terms with same replica indices
or coming from higher cumulants of the probability
distribution will produce, apart from
trivial or irrelevant contributions, a term of the form:
$$
\int d^2x\displaystyle\sum_{i=1}^{n} \left(\Phi_{1}^i(x) +
\Phi_{2}^i(x)\right) 
$$
Here we have denoted by $\Phi$ the $\phi_{3,1}$ operator. 
The problem of one $M_p$ model perturbed by the $\phi_{3,1}$ operator has
been studied extensively in \cite{zam}. There is a non trivial infra-red (IR)
fixed point, and it has been shown that the system flows to the $M_{p-1}$
model. This fixed point will also be present in the 
(RG) behavior of our particular model and will be of particular interest in
the quenched case.

\section{Generic case}
We first consider the case where we have $n$ replicas and thus our model
consists in $2n$ minimal models 
coupled together.
The idea is to study the (RG) behavior of
(\ref{qham}) for $n>2$ and
to identify the different fixed points that can appear.
As explained in the previous section, the model that we consider in the
following is described by the more general Hamiltonian~:
\beq
\label{ham}
H=\displaystyle\sum_{i=1}^{n} H_{0,1}^i + H_{0,2}^i
+ \lambda \int d^2x \displaystyle\sum_{i=1}^{n}\left( \Phi_{1}^i(x) +
\Phi_{2}^i(x)\right) +
g \int d^2x \displaystyle\sum_{i\neq j}^{n} \left(\ve_{1}^{i} \ve_{2}^{i}
\ve_{1}^{j} \ve_{2}^{j}\right)(x)
\eeq
where $ H_{0,1}^i$ and $ H_{0,2}^i$ are the Hamiltonians of the unperturbed
systems, each of them corresponding to a minimal model  $M_{p}$. Note that we
have replaced $\sigma$ by $-g$, thus the physical case for a random model
corresponds to $g<0$.  The 1-loop RG
equations can be easily obtained from the operator algebra of the
perturbing fields~:
$$
\displaystyle\sum_{i\neq j}^{n} \left(\ve_{1}^{i} \ve_{2}^{i}
\ve_{1}^{j} \ve_{2}^{j}\right)(x) ~~
\displaystyle\sum_{k\neq l}^{n} \left(\ve_{1}^{k} \ve_{2}^{k}
\ve_{1}^{l} \ve_{2}^{l}\right)(y) ~~\rightarrow ~~
4 (n-2) |x-y|^{-2+2\ep}
\displaystyle\sum_{i\neq j}^{n} \left(\ve_{1}^{i} \ve_{2}^{i}
\ve_{1}^{j} \ve_{2}^{j}\right)(y)
$$
$$
+ 4 (n-1) C_{\ve \ve}^{\Phi} |x-y|^{-2+{8\ep\over 3}} 
\displaystyle\sum_{i=1}^{n}\left( \Phi_{1}^i(y) +
\Phi_{2}^i(y)\right)
+\cdots
$$
$$
\displaystyle\sum_{l=1}^{n}\left( \Phi_{1}^l(x) +
\Phi_{2}^l(x)\right) ~~
\displaystyle\sum_{i=1}^{n}\left( \Phi_{1}^i(y) +
\Phi_{2}^i(y)\right) ~~\rightarrow ~~
 C_{\Phi \Phi}^{\Phi}  |x-y|^{-2+{4\ep\over 3}} 
\displaystyle\sum_{i=1}^{n}\left( \Phi_{1}^i(y) +
\Phi_{2}^i(y)\right)
+\cdots
$$
\beq
\label{oa}
\displaystyle\sum_{l=1}^{n}\left( \Phi_{1}^l(x) +
\Phi_{2}^l(x)\right) ~~
\displaystyle\sum_{i\neq j}^{n} \left(\ve_{1}^{i} \ve_{2}^{i}
\ve_{1}^{j} \ve_{2}^{j}\right)(y) ~~\rightarrow ~~
 4 C_{\Phi \ve}^{\ve}  |x-y|^{-2+{4\ep\over 3}}
\displaystyle\sum_{i\neq j}^{n} \left(\ve_{1}^{i} \ve_{2}^{i}
\ve_{1}^{j} \ve_{2}^{j}\right)(y)+\cdots
\eeq
where we have omitted the descendent terms. $C_{\Phi \Phi}^\Phi$ and $
C_{\Phi \ve}^{\ve}$ 
are the structure constants of the model $M_p$. They are symmetric under
permutation of the three indices and their values can be obtained from
\cite{df}:
$$
 C_{\Phi \ve}^{\ve} = {\sqrt{3}\over 2}~+~O(\ep)~~~;~~~ 
 C_{\Phi \Phi}^{\Phi} = {4\over \sqrt{3}}~+~O(\ep)
$$
Using the formula of the 1-loop RG equation \cite{dber}, \cite{zam}~:
$$
\dot{g}_i = (2-dim(g_i))g_i - \pi K_{jk}^i g_j g_k
$$
(where $g_i=g, \lambda$) and the values of the generalized structure
constants $K_{jk}^i$ obtained from eq. (\ref{oa}), we get the
following system of equations~: 
$$
\dot{g} = 2\ep g -  (n-2) g^2 -  \lambda g + \cdots
$$
\beq
\label{rg}
\dot{\lambda} = {4\ep \over 3}\lambda - {\lambda^2 \over 3} -   
{3\over 2} (n-1) g^2 + \cdots
\eeq
after the trivial redefinitions $ g \rightarrow g/4\pi ; \lambda
\rightarrow \lambda/ 4\pi \sqrt{3}$.
The first step in the study of the RG flow is to find all the fixed points of
eq. (\ref{rg}). This is easily done by solving eq. (\ref{rg}) with the
conditions $\dot{g}=\dot{\lambda}=0$. For $n>2$ we get 4 solutions~:
\beq
\label{fp1}
\lambda = 0 ~~~;~~~ g=0
\eeq
\beq
\label{fp2}
\lambda = 4 \ep  ~~~;~~~ g=0
\eeq
\bea
\label{fp3}
g = -{2\ep \sqrt{2} \over \sqrt{2(n-2)^2 +9(n-1)}}\quad &;& \quad \lambda
= 2\ep  (1+ {{\sqrt{2}(n-2)}\over \sqrt{2(n-2)^2 +9(n-1)}})\nn \\
\label{fp4}
g = {2\ep \sqrt{2} \over \sqrt{2(n-2)^2 +9(n-1)}}\quad &;& \quad \lambda
= 2\ep  (1- {{\sqrt{2}(n-2)}\over \sqrt{2(n-2)^2 +9(n-1)}})
\eea
Note that (\ref{fp1}) and (\ref{fp2}) are the two fixed points
present in the work of Zamolodchikov \cite{zam}, who found that a pure
$M_p$ model, perturbed by a $\phi_{3,1}$ operator flows to the point
(\ref{fp2}) which correspond to the $M_{p-1}$ model.
The next step is to study the stability of each of these fixed
points. This is done by linearizing (\ref{rg}) around the solutions given
above $g = g^* + \delta g ~;~ \lambda = \lambda^* +\delta  \lambda$ and
getting a linear system~:
$$
\left( {\delta \dot{g} \atop \delta  \dot{\lambda}}\right) = A
\left( {\delta g \atop \delta  \lambda}\right)  
$$
The eigenvalues of the matrix $A$ for each of the cases
(\ref{fp1}) to (\ref{fp4}) give us information about the stability of these
points. It is easy to see that for (\ref{fp1}) both eigenvalues are positives
indicating that this fixed point is unstable in all directions, while for
(\ref{fp2}) we have a stable fixed point in both directions. For
(\ref{fp3}) and (\ref{fp4}) we obtain one real negative and one real
positive  
eigenvalue, that is, these points are stable in one direction and unstable in
the other. Thus, they can be reached only if we fine tune the values of $g$
and $\lambda$ to keep our system in the stable line of these
points. Studying in detail the flow diagram of (\ref{rg}) we can see that the
initial conditions $\lambda = 0~;~g\neq 0$ will flow far from our fixed
points toward either a massive theory or another fixed point which can not
be seen at this order in perturbation theory.
\section{Quenched system}
We now turn to the case $n=0$ which correspond to the quenched system of
two randomly coupled minimal models. By just putting $n=0$ in (\ref{rg}) we
get our new RG equations:
\bea
\dot{g} &=& 2\ep g + 2 g^2 - \lambda g + \cdots \nn\\
\label{qrg}
\dot{\lambda} &=& {4\ep \over 3}\lambda - {\lambda^2 \over 3} +  
{3\over 2} g^2 + \cdots 
\eea
Solutions (\ref{fp1}) and (\ref{fp2}) are still valid with the same kind of
stability but now there is no more fixed point solutions with $g\neq
0$ (points (\ref{fp3}) and (\ref{fp4}) became complex.) So, (\ref{fp1}) and
(\ref{fp2}) are the only fixed points at 
this order 
in perturbation theory. Assuming that higher loop corrections to
(\ref{qrg}) will no change the qualitative behavior of the flow near our
two fixed points, we can see that a system with initial conditions $\lambda_0
= 0 ~;~ g_0 < 0$  will flow toward the point  (\ref{fp2}). This is
supported by the numerical calculation of the RG flow (\ref{qrg}) in figure
1 for different values of $g_0$. 
\[
\epsffile[100 100 300 450]{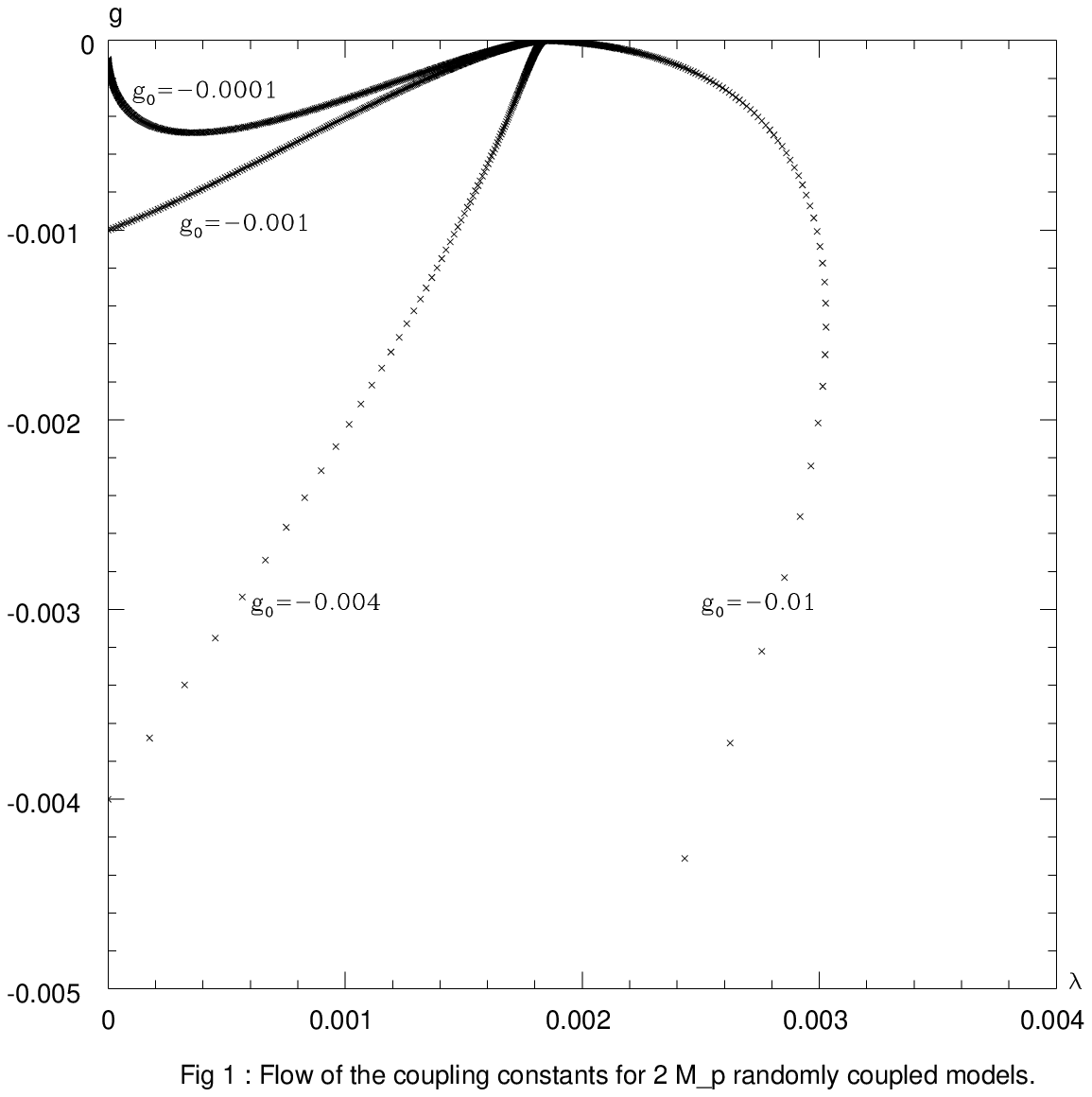}
\]
Initial conditions $\lambda_0 =0~;~g_0<0$ are precisely what we expect for
the case of a quenched system, since the term in the bare Hamiltonian
proportional to the $\Phi$ operator will have a factor $n$, coming from
the contraction of the other pairs of energy operators in
(\ref{qham}). Then, in the limit $n=0$, $\lambda_0$ cancels and $g_0=-\sigma
< 0$. 
So, adding a small random coupling 
trough the energy densities of two $M_p$ minimal model will drive our
system at large scales to two decoupled $M_{p-1}$ models.
\section{Generalization to $M$ coupled models}
In this section we will consider the generalization to $M$ coupled models.
The hamiltonian (\ref{ham}) is thus replaced by
\beq
\label{Nham}
H=\displaystyle{\sum_{a=1}^{M}\sum_{i=1}^{n} H_{0,a}^i }
+ \lambda \int d^2x \displaystyle\sum_{a=1}^{M}\sum_{i=1}^{n}\left(
\Phi_{a}^i(x) \right) +
\rho \int d^2x \displaystyle\sum_{a\neq b,c\neq d}^{M}\sum_{i\neq j}^{n}
\left(\ve_{a}^{i} \ve_{b}^{i} \ve_{c}^{j} \ve_{d}^{j}\right)(x)
\eeq
In the following computations, it will be more convenient to express the
last term like
\bea
\rho \int d^2x \displaystyle\sum_{a\neq b,c\neq d}^{M}\sum_{i\neq j}^{n}
\left(\ve_{a}^{i} \ve_{b}^{i} \ve_{c}^{j} \ve_{d}^{j}\right)(x)&=&\nn\\
g \int d^2x \displaystyle\sum_{a\neq b}^{M}\sum_{c\neq  d}^{M}\sum_{i\neq
j}^{n} 
\left(\ve_{a}^{i} \ve_{b}^{i} \ve_{c}^{j}
\ve_{d}^{j}\right)(x) 
&+&\sigma \int d^2x \displaystyle\sum_{<a, b, c, d>}^{M}\sum_{i=1}^{n}
\left(\ve_{a}^{i} \ve_{b}^{i} \ve_{c}^{i} \ve_{d}^{i}\right)(x)
\eea
Here $<a,b,c,d>$ means all the summation over $a, b, c, d$ which take
different values two by two. Under the renormalisation group
transformations, each of these terms is going to behave differently and
thus there will be two different equations for $g$ and $\sigma$. In fact
this is only true for $M\geq 4$. For $M < 4$, the $\sigma$ term is
absent. From the set of equations with $g, \sigma$ and $\lambda$, we will
still be able to recover the case $M=2$ and $M=3$ by simply suppressing
$\sigma$. The case $M=2$ was 
already considered in the previous section and the case $M=3$ will be
mentioned at the end of this section. The renormalisation group
equations for the parameters $g, \sigma$ and $\lambda$ are trivial to
compute by generalizing computations of previous sections. We obtain 
\bea
\label{Nrg}
\dot{g} &=& 2\ep g - {1\over 2}g^2(M(M-1)(n-2)+4(M-2)^2)  -
g\sigma(M-2)(M-3)-\lambda g + \cdots \nn \\ 
\dot{\sigma} &=& 2\ep \sigma - {3\over 2}\sigma^2(M-4)(M-5)-{3\over
2}g^2(n-1)M(M-1) - \lambda \sigma  + \cdots \\ 
\dot{\lambda} &=& {4\ep \over 3}\lambda - {1 \over 3}\lambda^2 -   
{3\over 4}g^2 (n-1)M(M-1)^2  -{1\over 4}\sigma^2(M-1)(M-2)(M-3)+ \cdots \nn
\eea
Going directly to the quenched case, we found
\bea
\label{NNrg}
\dot{g} &=& 2\ep g - g^2(2(M-2)^2-M(M-1))  -
g\sigma(M-2)(M-3)-\lambda g + \cdots \nn \\ 
\dot{\sigma} &=& 2\ep \sigma - {3\over 2}\sigma^2(M-4)(M-5)+{3\over
2}g^2M(M-1) - \lambda \sigma  + \cdots \\ 
\dot{\lambda} &=& {4\ep \over 3}\lambda - {1 \over 3}\lambda^2 + 
{3\over 4}g^2 M(M-1)^2  -{1\over 4}\sigma^2(M-1)(M-2)(M-3)+ \cdots \nn
\eea
The next step consists in computing the fixed points associated with these
equations. Using an $\ep$ expansion, we got the following points: first,
because $\dot{g}=g(...)$, one set of solutions is given by $g=0$ and, after
some additional computations
\bea
\sigma=0 \qquad &;& \qquad \lambda=0 \\
\label{s2}
\sigma=0 \qquad &;& \qquad  \lambda=4\ep \\
\sigma=\ep x \qquad &;& \qquad  \lambda=2\ep (1-y)\\
\sigma=-\ep x \qquad &;& \qquad  \lambda=2\ep (1+y)
\eea
where we have defined the following quantities
\bea
x&=&{4 \over \sqrt{3}} \left[(M-1)(M-2)(M-3)+3(M-4)^2(M-5)^2
\right]^{-{1\over2}}\nn\\   
y&=&{3x\over 4}(M-4)(M-5)\nn
\eea
In addition, there is a second set of solutions with $g\neq  0$ (with some
extra conditions on $M$, see bellow.) After some
more tedious computations, we get
\bea
\label{s5}
g=-\ep X_M^{+}\qquad &;& \qquad \sigma=-\ep Z^{+}_M X_M^{+}\\
\label{s6}
g=-\ep X_M^{-}\qquad &;& \qquad \sigma=-\ep Z^{-}_M X_M^{-}\\
\label{s7}
g=\ep X_M^{+}\qquad &;& \qquad \sigma=\ep Z^{+}_M X_M^{+}\\
\label{s8}
g=\ep X_M^{-}\qquad &;& \qquad \sigma=\ep Z^{-}_M X_M^{-}
\eea
and $\lambda=2\ep-g(2(M-2)^2-M(M-1))-\sigma(M-2)(M-3)$. We also used the
following definitions 
\bea
Z_M^{\pm}&=&{-(2(M-2)^2-M(M-1))\pm 2 \sqrt{f_M} \over
2((M-2)(M-3)-{3\over2}(M-4)(M-5))}\nn\\
X_M^{\pm}&=&{\sqrt{3}\over 2} ({1\over
3}(Z_M^{\pm}(M-2)(M-3)+(2(M-2)^2-M(M-1)))^2\\
&&+ {1\over
4}(Z_M^{\pm})^2(M-1)(M-2)(M-3)-{3\over 4}M(M-1)^2 )^{-{1\over2}}\nn\\
f_M&=&M^4-17M^3+65M^2-64M+16\nn
\eea 
These last four solutions do not exist for every $M$. We have some
additional constraints:  $f_M$ is positive, and the 
fixed points real, only for $M \geq 13$; $X_M^{+}$ is real only for $M \leq
66$. Thus the four solutions (\ref{s5}-\ref{s8}) are real for $13 \leq M \leq
66$ while for $M \geq 67$ only solutions (\ref{s6},\ref{s8}) do exist.
The next step is to study the stability of these solutions. Let first note
that the physical initial conditions are 
\beq
\lambda=0 \qquad ; \qquad g=\sigma < 0
\eeq
Because $\dot{g}=g(...)$, we have the condition $g\leq 0$. Then, the
last two solutions (\ref{s7}) and (\ref{s8}) can be immediately
discarded. Thus only 6 solutions can be relevant. The first one
($g=\sigma=\lambda=0$) is obviously unstable and can also be
discarded. The second one ($g=\sigma=0, \lambda=4\ep$) turns out to be
stable ($\delta \dot{g}=-2\ep \delta g, \delta \dot{\sigma}=-2\ep \delta
\sigma, \delta \dot{\lambda}=-{4\over 3}\ep \delta \lambda$ ) and this
independently of $M$. This point is just the stable fixed point 
(\ref{fp2}) of the previous section. The next two solutions ($g= 0, \sigma
=\pm \ep x, \lambda= 2\ep (1\mp y)$) are also
unstable. This can be seen by noticing that 
\beq
\left( {\delta \dot{\sigma} \atop \delta  \dot{\lambda}}\right) = \ep A_{\pm}
\left( {\delta \sigma \atop \delta  \lambda}\right)  
\eeq
with 
\bea
A_{\pm} = \mp {1\over 2} x \left( \begin{array}{cc}
3(M-4)(M-5) & 2 \\
(M-1)(M-2)(M-3) & -2(M-4)(M-5) \end{array}\right)
\eea
and then 
\beq
\det{A_{\pm}} =-{1\over 4}x^2 (6(M-4)^2(M-5)^2+2(M-1)(M-2)(M-3)) < 0
\eeq
for all $M\geq 4$, which implies that there is two eigenvalues with opposite
sign. Then we remain with the last two solutions, (\ref{s5}) and
(\ref{s6}) for the previously mentioned values of $M$. Again these two
solutions will be unstable. Here, it is not 
possible to give an analytical proof for all $M$. So we computed
numerically the eigenvalues of the matrix $A_{\pm}$ defined by
\bea
\left(\begin{array}{cc} 
\delta \dot{g} \\ \delta \dot{\sigma} \\ \delta \dot{\lambda} 
\end{array} \right) 
= 
\ep A_{\pm} \left(\begin{array}{cc}
\delta{g} \\ \delta \sigma \\ \delta  \lambda
\end{array} \right)  
\eea
We found that for each $13\leq M \leq 1000$, at least one of the
eigenvalue of $A_{-}$ is positive, the same being true for $A_{+}$ for each
$13\leq M \leq 66$. 

Finally the last step is to check if under the renormalisation group, we
are going to reach the only remaining fixed point, (\ref{s2}). Again this
was done by constructing numerically the flow diagram. We construct it for
every value of $M$ between $4$ and $1000$ and for each of these flows,
we found that we reach the stable fixed point (\ref{s2}).

Before going to the discussion, let's also mention the case
$M=3$. Renormalisation group equations for this case are obtained from
(\ref{NNrg}) by suppressing the $\sigma$ terms. We
found $2$ solutions to these equations: the trivially unstable solution
($g=\lambda=0$) and the stable solution ($g=0, \lambda=4\ep$). Again, we
remain with only one solution which corresponds to $3$ decoupled $M_{p-1}$
models.

\section{Discussion}

In this paper, we have considered a very particular way of adding randomness
for more general systems than the well studied Ising or Potts models with
random bonds. We considered the case of two minimal $M_p$ models randomly
coupled and our 
1-loop calculation shows that (weak) randomness can be easily chosen such
that criticality is maintained and our system will behave at large
distances as two decoupled $M_{p-1}$ models. Then we generalized the study to
$M$ randomly coupled $M_p$. Again we found that at the
1-loop order, these models will behave at large distance like $M$ decoupled
$M_{p-1}$ models. In fact, the operator algebra of the perturbation induced
by randomness produce in the effective action 
a supplementary term which drift our system to a unitary I.R. fixed
point. These results seem to tell us
that the critical behavior of some two dimensional systems in the presence
of randomness depend crucially on the particular model and the kind of
randomness we are considering, in contrast to what seems to happens in the
cases studied in \cite{cardy,sent,domany}. We expect that higher loop
corrections in our 
renormalization group equations shouldn't modify the qualitative behavior
of the flow. However, an analytic solution for the coupling flow should
make more concrete these conclusion.

\end{document}